# Cosmic-ray reactions in the atmosphere and in nuclear emulsions: Radiocarbon dating and "disintegration stars"

Brigitte Strohmaier, University of Vienna, Faculty of Physics – Nuclear Physics, Vienna, Austria

## Abstract

Cosmic rays were discovered by Victor Hess in 1912 through their effect on the conductivity of the atmosphere. It took several decades of research to understand their nature and composition. The photographic method of particle detection, developed and refined by Marietta Blau, played an important part in making visible cosmic-ray reactions in nuclear emulsions as "disintegration stars" in 1937. In the high atmosphere, cosmic-ray reactions produce not only charged particles, but also neutrons which lead to the formation of radiocarbon. The dating technique based on $^{14}$C was developed in 1946 by Willard Libby. The present work describes how members of the Institute for Radium Research in Vienna contributed to the discovery of cosmic rays, "disintegration stars" as well as the application of radiocarbon dating.

## 1. Radiocarbon dating and cosmic rays

Age determination by means of radiocarbon is based on the constant presence of $^{14}$C relative to stable carbon isotopes in the atmosphere and hence in all living organic material. When the plant or animal dies the intake of carbon ends and the amount of $^{14}$C decreases due to radioactive decay. This occurs via β-emission with a half-life of 5,730±40 years. Sensitive radiation detectors enable the determination of the $^{14}$C content which serves as a measure for the time passed since the end of ingestion of carbon.

The pioneer of this method, Willard Libby (1908–1980; Fig. 1), had earned his bachelor and PhD at UC Berkeley in the early 1930s with studies of radioactive elements and measurements of weak natural and artificial radioactivity. He entered the faculty of UC Berkeley, but as of 1942 got involved in the Manhattan Project, working at Columbia University on various types of processes for uranium enrichment.[1]

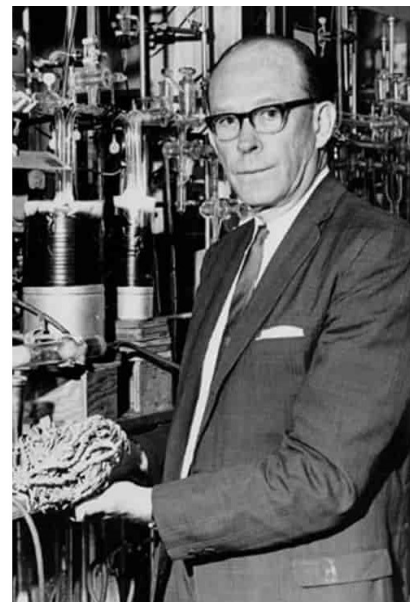

Fig. 1: Willard Libby in the Laboratory, 1960s (© Ole Bostrup 2022)

After World War II, Libby became a professor at the University of Chicago where in the Chemistry Department of the new Institute for Nuclear Studies he resumed his radioactivity studies.[1] He could build on two discoveries: that of $^{14}$C by Martin Kamen and Sam Ruben[2] at the UC Berkeley Radiation Lab in 1940, and that of the production of neutrons by cosmic rays in the upper atmosphere by Serge A. Korff[3] in 1939. These neutrons react with nitrogen in the atmosphere to form radiocarbon: n + $^{14}$N → $^{14}$C + p.

Whereas it was extremely difficult to predict the types of nuclei that might be produced by the GeV primary cosmic rays, the neutrons being secondaries were in the MeV energy range. Korff had noted that the principal effect of neutrons in air would be the formation of radio-carbon when impacting nitrogen – oxygen is essentially inert to neutrons – and this, at a rate equal to that at which it decayed to $^{14}$N by β-decay ($^{14}\text{C} \rightarrow {}^{14}\text{N} + \text{e}^- + \bar{\nu}$) so that a steady-state condition should have established, the above mentioned presupposition of the dating method.

The $^{14}$C dating method was published by Libby theoretically in 1946 and in a monograph in 1955, and awarded the Nobel Prize in Chemistry in 1960. Still in his Nobel Lecture, Libby states that "ignorance of billion-electron-volt nuclear physics (cosmic ray energies are in this range) was so abysmal at the time and incidentally fourteen years later still is so abysmal, that it is nearly impossible to predict with any certainty the effects of the collisions of the multibillion-volt primary cosmic radiation with air."

Let us now turn to those scientists who dared look into that abyss.

## 2. The discovery of cosmic rays

At the end of the 18$^{th}$ century, it was observed that a charged object would eventually discharge despite insulating suspension, and concluded that the air was conductive.

At the beginning of the 20$^{th}$ century, ions, x-rays and particles of radioactive decay were already known and detected by ion chambers and leaf electroscopes. The atmosphere's non-vanishing conductivity was thought to be due to ionization by energetic radiation, possibly from radioactive material in the earth's crust. In order to test whether the ionization (number of ions per cm$^3$ and sec) decreased with increasing distance from the surface, Theodor Wulf, a German Jesuit who taught at an order college in the Netherlands, measured the ionization at locations above ground, including the Eiffel Tower, in 1909, and Karl Bergwitz (Braunschweig, 1908) came up with balloon rides for such measurements. Albert Gockel, a German physicist teaching at Freiburg, Switzerland, undertook balloon flights to altitudes of up to 4500 m in 1909 and 1910. He recognized a slight increase of ionization with increasing altitude, his finding being subject to some doubt, though.[4]

### 2.1. The Institute for Radium Research in Vienna

In 1911, therefore, Viktor Franz Hess turned to this problem. At that time, he was First Assistant at the Institute for Radium Research in Vienna. Physics institutes had existed at the University of Vienna since 1848, and radioactivity was investigated there, mainly by Stefan Meyer (Fig. 2), as early as 1899, only three years after its discovery by Henri Becquerel. Two years later the Imperial Academy of Sciences established a commission to investigate radioactive substances and had radium extracted from ten tons of tailings from the uranium production in Joachimsthal in Bohemia (at that time part of the Austro-Hungarian Empire). The Academy therefore owned the largest quantity of radium in the world. Because of the inadequate technical equipment at the old physics institute, the true value of this resource could not be exploited. In 1908, Karl Kupelwieser, a lawyer, made a donation to the Imperial Academy of Sciences for the construction and equipping of a building devoted to the physical investigation of radium. It was dedicated in October 1910 (Fig. 3) and was called Radium Institute for short from the beginning. Stefan Meyer had planned the building and equipment of the Vienna Radium Institute (VRI) and was, therefore, considered its creator. He was the head of the institute. The staff of the Radium Institute was employed partly by the Academy of Sciences and partly by the University of Vienna.

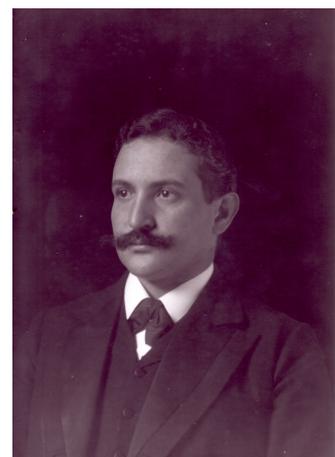

Fig. 2: Stefan Meyer, 1908 (Austr. Central Library for Physics)

## 2.2. Viktor Franz Hess

Viktor Hess (1883–1964; Fig. 4) was born in Styria and studied physics at the University of Graz. After his PhD graduation under the auspices of Emperor Francis Joseph in 1906, he went to the University of Vienna and in 1910 was employed at the new Radium Institute. The same year, he acquired the right of teaching with a work on *Absolute Determination of the Atmosphere's Contents of Radium Progeny*.

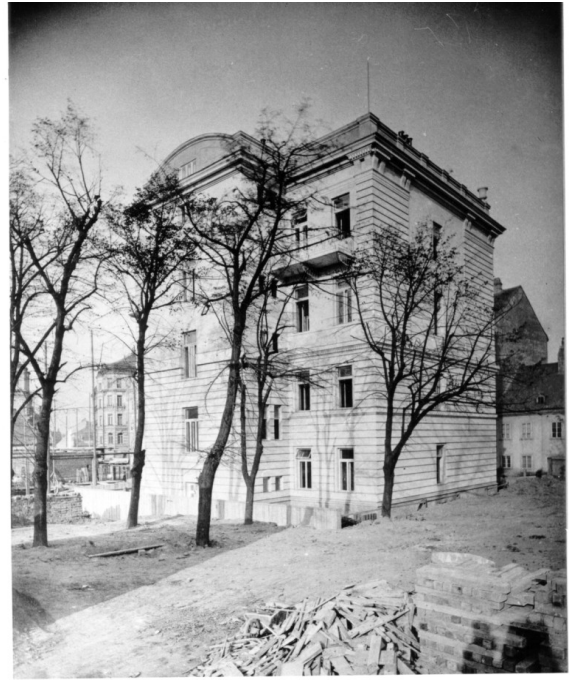

Fig. 3: The Vienna Radium Institute in 1910, view from the rear courtyard (Austr. Acad. Sci., Photo Archive A-0142-C)

Hess took up the matter of balloon flights, using the radiation apparatus designed by Wulf with an airtight ionization volume and a wall thickness of 2 mm to withstand air-pressure variations and let pass γ-radiation only. For his flights, he used free balloons, filled with coal gas ($\geq$ 50% $H_2$) on all but one flights and pure hydrogen on the last one. The only means of control was releasing gas and reducing ballast. The balloons were supplied by the Austrian Aero Club whose members belonged to the military aeronautic institution of the Austrian-Hungarian army. On two balloon rides in 1911 and seven in 1912 which carried him as far as Bohemia, he found that the ionization was constant up to an altitude of 1000 m and increased above, indicating that in addition to the substances in the earth's crust some penetrating radiation must ionize the air with effectiveness increasing with increasing altitude. On the flights in 1912, some of which were carried out during night or a solar eclipse, he applied two Wulf γ-detectors and a β-electrometer at the same time, going as high as 5,200 m. The finding was that β- and γ-measurements showed the same dependence on altitude, but parallel γ-detection sometimes yielded different ionization at the same altitude. No influence of the sun was observed. The results were explained by assuming a radiation of very high penetrability which enters the atmosphere from above and even in its lowest layers causes part of the effects detected in closed chambers. The intensity of this radiation seemed to vary in time. The sun could not be considered the direct source of this hypothetical penetrating radiation, as long as one thought of a direct γ-radiation with linear propagation.[5] Consequently, the radiation was named ultra-γ-radiation, penetrating radiation, Hess's radiation or cosmic radiation; in German also the term *Höhenstrahlung* was used.

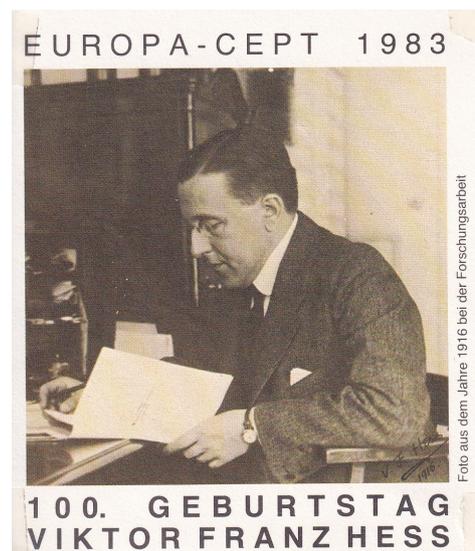

Fig. 4: Viktor Hess at the Vienna Radium Institute, 1916 (First-day cover of a stamp on the occasion of his 100[th] birthday)

# 3. The identification of cosmic rays
## 3.1. Further electric measurements

As with a maximum altitude of over 5 km, Hess had reached the limit the human organism can safely withstand in an open gondola, the Prussian physicist Werner Kolhörster, faculty at the University of Halle, sent unmanned balloons equipped with ionization chambers up to 9 km, finding a further increase in ionization and hence confirming Hess's extraterrestrial cosmic radiation.

World War I and the subsequent economic problems interrupted the experiments. Hess became professor in Graz (Styria) in 1920 and in Innsbruck (Tyrol) in 1931, and installed observatories for permanent cosmic-ray registration at mountains (Hochobir, 2139 m; Hafelekar, 2300 m). Kolhörster, professor for radiation physics in Berlin since 1935, founded an Institute for Cosmic-Ray Research. Together with Walther Bothe, he intended to verify cosmic rays as energetic γ-radiation. They placed metal plates (iron, lead) as absorbers between two Geiger counters, as a γ-quantum can be detected in a Geiger counter only via an electron hit out of a neutral atom. They found coincidences, i. e., simultaneous events in both counters, indicating that the ejected electron had crossed them both. When they tried to estimate the γ-energy by increasing the absorber thickness until the coincidences be suppressed, they found this impossible in 75% of the cases, and concluded that cosmic rays are not γ-radiation, but charged particles of high penetrability. They could prove that also the secondary radiation produced by the interaction of primary cosmic rays with the atmosphere consists of charged particles.[6]

Further balloon flights with ionization-chamber measurements were performed in the 1930s. Studies of the cosmic ray intensity as a function of altitude were expected to give evidence regarding the nature of cosmic rays as the variation should follow different laws according as the rays are electrons or photons.[7] Measurements of the relative intensity of cosmic rays over different parts of the earth were designed to show any effect due to the earth's magnetic field.[7]

## 3.2. Detector development at the VRI in the 1920[s]

A new incentive to investigate methods of particle detection was triggered when scientists, also at the VRI, started studying "atomic disintegration", processes induced by α-particles from radioactive decay on stable nuclei, similar to the "transmutation" $^{14}N(\alpha,p)^{17}O$ first published by Ernest Rutherford in 1919. To prove such reactions, the fragments needed to be registered. The scintillation method in use for that purpose required the observation of light flashes caused by ionizing radiation on ZnS screens under the microscope in a dark chamber which was tiring and subject to error. Improved methods for fast-proton detection were desired. At the Radium Institute, such efforts were, on one hand, put forth by Berta Karlik who studied the scintillation process itself in order to possibly come up with a more objective way of registering the scintillation light. On the other hand, Marietta Blau[8,9] dedicated her further work to the technique of photographic detection of particles. Alternative methods of particle counting comprised various electrical techniques, like modifications of ionization chambers, and Wilson's cloud chamber, which was not available at the VRI before 1925/26.

## 3.3. The photographic detection of electromagnetic and particle radiation

In a photographic emulsion, small crystals of silver bromide (AgBr) are distributed in a transparent matrix of gelatin. The emulsion is coated on a thin glass plate. In the crystals, the elements are arranged as ions in a lattice ($Ag^+$, $Br^-$). All silver halides are sensitive to electromagnetic radiation and ionizing particles in that electrons produced by the radiation neutralize silver ions to form atoms which accumulate on silver development seeds existing in the emulsion. The resulting silver clusters form a latent image which is stable over long time and can be turned into a visible image by developing. The developer, a reducing agent, renders electrons to the silver ions which are reduced to metallic (visible) silver, preferably at the clusters of the latent image produced by exposure to light or particle radiation. In a fixing bath the unexposed parts of the layer are washed out.

Photography was applied for imaging of x-rays shortly after their discovery in 1895. The photographic effect of uranium radiation had led to the discovery of radioactivity, hence it was straightforward to apply photographic emulsions to register the tracks of α-particles from radioactive decay. Such studies were performed at the VRI by Wilhelm Michl in 1914.[10] He found that α-particles blacken the emulsion at discrete places, i. e., produce sequences of black dots (silver grains) at grazing incidence. After Michl had died in World War I, the matter was not further pursued.

### 3.4. The development of the photographic method at the VRI

Blau's task as of 1924 was to investigate whether fast protons emitted in (α,p) reactions could be observed through their photographic effect. It was clear that if photographs of particle paths could be taken, the tracks would remain stored on the plates and enable subsequent analysis, in contrast to the quickly vanishing scintillations.

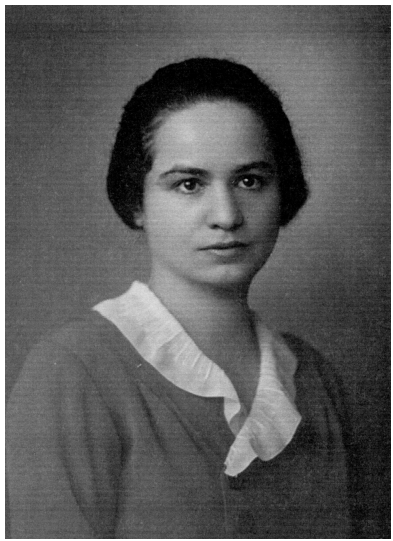

Fig. 5: Marietta Blau around 1927 (Courtesy Eva Connors-Blau)

Marietta Blau[8,9] (1894–1970, Fig. 5) obtained her general certificate of education (*Matura*) from the high school for girls run by the Association for the Extended Education of Women (*Verein für erweiterte Frauenbildung*) in 1914 and enrolled at the University of Vienna as a regular student of physics and mathematics the same year. With a doctoral thesis on a radiological topic, *The Absorption of Divergent γ-Rays*, Blau obtained her PhD in March 1919. After having spent several months as an observer with Guido Holzknecht at the Central X-ray Institute of the Vienna General Hospital, Marietta Blau was employed as a physicist at an x-ray tube factory in Berlin in 1921 and as assistant professor at the Institute for the Physical Bases of Medicine at the University of Frankfurt/Main in 1922. When she returned to Vienna the following year, she worked at the Radium Institute without pay, which was quite common for scientists at the VRI.

Blau constructed a high-yield proton source and experimented with various types of photographic plates, the essential properties being grain size, i. e., average size of AgBr crystals in the emulsion, and the layer thickness of the emulsion. Development conditions, in particular duration and temperature of development as well as composition and concentration of the developer, had to be adjusted to the features of the emulsion.

In 1927, Blau succeeded in detecting protons[11] whose tracks in the photographic layer consist of rows of black dots which correspond to developed grains. The length of a row is a measure for the range of the particles in the emulsion, i. e., the mean distance the particle travels until it is stopped due to successive energy losses, and hence its total energy. In order to obtain the particle energy, therefore, the total track length had to be recorded which due to the thinness of the emulsion layers (≤ 200 μm) was only achieved if the protons passed the emulsion approximately parallel to the layer surface. Using thicker emulsions required appropriate development conditions to enable a homogeneous development of the complete emulsion layer.

Beside proving the disintegrability of aluminum by α-particles, $^{27}Al(α,p)^{30}Si$, Blau evaluated thousands of α- and proton tracks with regard to length and number of blackened grains with the microscope in order to obtain quantitative data on the photographic effect of these particles.[12] She discussed the mechanism of the photographic effect itself and studied the fading of the latent image, i.e., whether and how fast the invisible image consisting of activated silver ions vanishes after exposure. Information about fading is necessary if the photographic plates are stored for some time between exposure and development, and especially for low-level measurements with long exposure times.

The visibility of proton tracks in the emulsion was often insufficient, though, and at the same time, competing radiation such as β- and γ-rays as well as visible light was registered on the photographic plates and impeded their evaluation with regard to particle tracks. Therefore, Blau considered the pre-treatment of emulsions by certain chemicals.

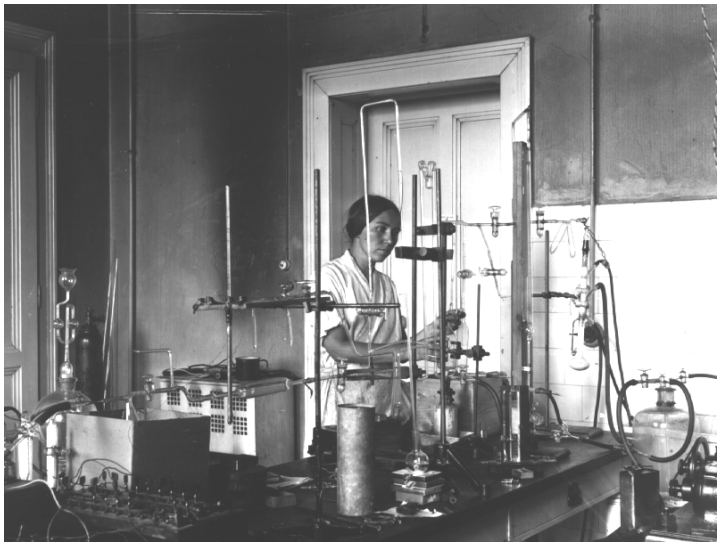

Fig. 6: Hertha Wambacher (Archive R. & L. Sexl)

Investigating how such impregnation of photographic plates affected their response to particle radiation was transferred on Hertha Wambacher (1903–1950; Fig. 6) in 1928. Blau supervised her doctoral thesis.[13] It turned out that the pre-treatment of the emulsions, mainly with dyes, not only suppressed the effect of background radiation, but also improved α-particle tracks and caused fast protons which had not been detectable at all, to yield observable tracks. Wambacher graduated in 1932. Blau and Wambacher worked together for six more years on methodical investigations of the photographic method, examining a large number of substances for impregnating photographic plates and providing an explanation of the effect of a desensitizer (with regard to visible light) as sensitizer for particle radiation.

## 3.5. Cosmic rays: fast protons and spallation stars

As soon as the photographic method was established as a tool for the detection of charged particles, Marietta Blau and her collaborator Hertha Wambacher turned to applying it to the exploration of cosmic rays. In 1932, twenty years after the discovery, conclusions on the nature of cosmic rays had been derived from ionization measurements at stations at different latitudes and from combined balloon and mountain data. Both the geographical distribution and the intensity as function of altitude were consistent with the assumption of charged particles as constituents of cosmic rays, as was the Bothe-Kolhörster experiment. But there was no way of reconciling the data with the hypothesis that primary cosmic rays consist of photons.[7]

The fact that photographic plates accumulate events over long periods of time made them particularly suited to detect the rare cosmic-ray hits. The knowledge of fading of the latent image turned out to be most valuable in cosmic-ray research with exposures lasting several months. In order to register as large a fraction of ionization tracks of high-energy particles as possible, increased thickness of the emulsion layers in the photographic plates was considered indispensable. New developing techniques were needed for the increased layer thickness.

In 1937 Blau and Wambacher approached Viktor Hess, the discoverer of cosmic rays, who in 1936 had been awarded the Nobel Prize in Physics for his pioneering work. In his observatory at Hafelekar, a 2300-meter mountain north of Innsbruck, the intensity of cosmic rays was constantly recorded, initially with ionization chambers. Upon their request to Hess, Blau and Wambacher were allowed to expose photographic plates at the cosmic-ray observatory.

On the plates exposed for four months at Hafelekar, many long tracks were found and assumed to be those of protons, either from recoil by neutrons or from atomic-disintegration processes, but not primary particles themselves, as no strong preference of vertical incidence was observed. Numerous tracks corresponded to ranges in air of over one and up to twelve meters. Due to these large ranges, most tracks did not end in the emulsion, hence the observed lengths did not equal the ranges (energies) of the particles. Therefore, these were determined from the distances of developed grains in the tracks. To enable such an evaluation, Blau and Wambacher analyzed a large number of tracks and derived a (preliminary) relation between grain distance and particle energy, ending up with an average particle energy of 12 MeV.

Even more importantly, a new pattern of tracks was discovered, namely that of several reaction products starting where a cosmic-ray induced nuclear reaction had taken place. Due to the starlike shape of these tracks (Fig. 7), they were called disintegration stars (*Zertrümmerungssterne*). From a pronounced center, up to twelve tracks proceed, varying in grain density and length. For several of sixty registered stars of "multiple disintegration" (spallation), Blau and Wambacher estimated the total energy of emitted particles to be several hundreds of MeV.

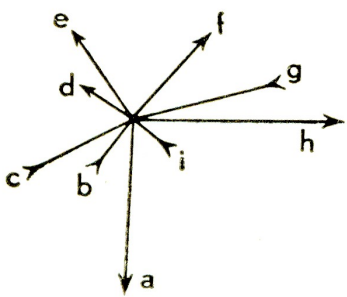

Fig. 7: Schematic drawing of a disintegration star (*Sitzungsber. Akad. Wiss. Wien, Math.-Nat. Kl.* IIa 146 (1937) 623)

Wambacher was the ambitious one who urged to publish the results[14] as soon as possible in order to keep others from getting in front of them. In September 1937 Blau made a request of Friedrich Paneth who had worked at the Radium Institute for several years before 1920 and was now professor at the Imperial College in London, whether they might include packets of photographic plates on balloon flights into the stratosphere in order to register particle tracks of cosmic rays in the emulsions. What followed was a lively correspondence on feasibility, achievable altitude, weight and wrapping of the photographic-plate packets, and optimization of the emulsions. Blau and Wambacher even received a grant from the Academy of Sciences enabling balloon flights with the emulsions. All this was interrupted, however, by the political events in Austria in 1938.

Marietta Blau left Vienna for a research stay at Oslo University on March 12, 1938, the evening before the Nazi invasion of Austria, one of the last Austrians to pass the German border. When meeting the German troops on her trip, she realized that – due to her Jewish descent – probably she would not be able to return from Oslo to Vienna.

Blau and Wambacher's discovery of the disintegration stars in 1937 was met with great interest in scientific circles. The theory of nuclear forces was just at the very beginning of its development. Werner Heisenberg had worked out a theory about cosmic-ray particles colliding with atomic nuclei leading to multi-particle emission. Blau and Wambacher's detection of such events now provided experimental evidence for his ideas and furthered the discussion of such processes.

After Blau's departure, Wambacher, an ardent Nazi, published on *Multiple Disintegration of Atomic Nuclei by Cosmic Rays* already in September 1938, and acquired her *Habilitation* (right of teaching at universities) on the basis of the work *Nuclear Disintegration by Cosmic Rays in Photographic Emulsions* in 1940. After the War, she was removed from the university like all members of the Nazi party, but had been deported to the USSR before. She returned in 1946. When she died of cancer in 1950, in several obituaries she was praised for her work as if she had been the preeminent author of the development of the photographic method and the discovery of the disintegration stars. Blau's leading role in this team is insufficiently expressed.

Blau, on the other hand, emigrated first to Mexico City, later to the U.S. In 1950 she was employed at BNL where she adapted the photographic method for investigating the interaction of high-energy particles with emulsions. Now that she had access to the most modern and expensive research techniques, she not only had to cope with Wambacher being given all credit for their joint work and success, but also with the fact that the Nobel Prize in Physics was awarded to Cecil F. Powell for the development of the photographic method for particle detection. Powell had started to work on the photographic method at Bristol University in 1938, taking up and developing further the work Blau had completed up to the point when she was forced to leave Vienna. In 1947, he discovered tracks of π-mesons in photographic plates exposed to cosmic rays, which was partly the reason he was awarded the Nobel Prize. Blau's pioneer work went unrewarded. In fact, Erwin Schrödinger had proposed Blau and Wambacher for the Nobel Prize in Physics in 1950, pointing out how important their method was in exploring cosmic rays in general and that, in particular they had been the first to interpret the stars correctly as atomic disintegration induced by cosmic rays. The Nobel Prize nomination by Schrödinger was unsuccessful.

Blau moved to the University of Miami in Coral Gables in 1956, where she set up a research program in particle physics in collaboration with Arnold Perlmutter (1928–2017). When she returned to Vienna in 1960, she functioned as advisor to a high-energy group which analyzed photographic plates as well as bubble-chamber photographs taken at the European Organization for Nuclear Research, CERN, near Geneva. At the end of 1964, Blau quit her activities at the Radium Institute due to health trouble. She died in January 1970 of lung cancer.

## 4. Radiocarbon dating at the VRI
### 4.1. Low-level measurement of β-radioactivity

After World War II, Berta Karlik (1904–1990; Fig. 8) became director of the Vienna Radium Institute. At a time when in the U.S. science flourished due to the use of particle accelerators and nuclear reactors, the whole of Austria as well as its capital Vienna suffered from severe war damage and were divided into four zones occupied by the victorious allies USA, Great Britain, France and USSR until 1955. There was no funding or material for scientific equipment. Karlik knew that the only way of succeeding in nuclear-physics research under such circumstances was to count on know-how rather than costly facilities. The concept of her directorship comprised three items which could be realized in a relatively inexpensive way by manufacturing the required electronic devices in the institute's workshop: The construction of a Cockcroft-Walton accelerator for deuterons to produce mono energetic neutrons for nuclear-reaction studies; the development of a high-performance γ-spectrometry system, and the installation of a facility for radiocarbon dating.

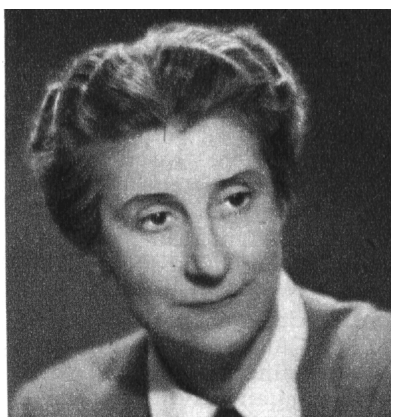

Fig. 8: Berta Karlik
(Archive R. & L. Sexl)

The latter task was transferred on Heinz Felber (1922–2013; Fig. 9) who had graduated from high school and started studying chemistry during World War II, but was called up to the German air force in 1941. He fell prisoner of war and returned in 1946, when he continued studying chemistry, but soon switched to physics. He worked his dissertation[15] at the VRI and obtained his PhD in spring of 1951. There was no position vacant at the VRI then. Therefore, he assumed one in industry, specifying on electrotechnical research. In fall of 1955 he was employed at the VRI and from 1958 to 1962 set up the apparatus for $^{14}$C dating together with Peter Vychytil.[16]

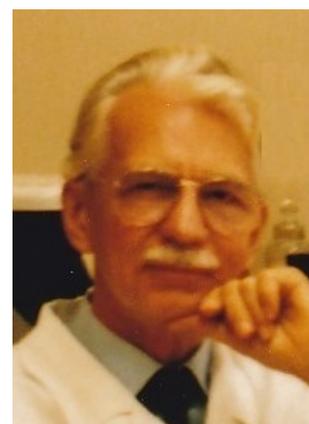

Fig. 9: Heinz Felber, 1986
(Courtesy Edwin Pak)

At this time, it was well established that primary cosmic rays consist of positively charged atomic nuclei, among which protons form the largest portion. Nuclei of all elements up to the heaviest are present as well; their intensity decreases rapidly as the nuclear charge number increases. Primary cosmic rays induce nuclear reactions when reaching the earth's atmosphere. Short-lived particles (mesons, hyperons) as well as γ-rays, electrons, and positrons are produced in these reactions. As further interaction occurs, particle cascades and showers take place. A schematic drawing[17] depicts these processes (Fig. 10).

The application of the radiocarbon method was consuming with regard to apparatus, despite its simple principle. The energy maximum of the $^{14}$C β-spectrum is 156 keV only. Radiation that low in energy would not leave a solid sample; therefore, the carbon of the sample material was transformed into methane which was used as counting gas of a proportional counter. Another problem was posed by the low specific activity of $^{14}$C: In atmospheric carbon and hence in recent organic matter, only one out of $10^{12}$ carbon atoms is $^{14}$C. In case of old samples, intensities of 0.1 decays per minute had to be counted which were superimposed by perturbations exceeding the effect by three orders of magnitude. Their origin was environmental radiation: Construction material as well as air of buildings contains radon. But in the Radium Institute of the 1960$^s$ the background level exceeded the common values by far. In decades of radium research, chemical work had been performed using open emanating substances, radon giving rise to radioactive progeny. Also for decades, barrels of tailings from the Joachimsthal radium extraction had been stored in the cellar. Later, for the operation of the neutron generator targets were loaded with tritium.[18] Another essential source of background is of extraterrestrial origin: cosmic rays. To reduce the background level, the counter was coated with a 20-cm thick iron shield weighing two tons. For this screen, old iron from times before the industrial application of radioactive isotopes had to be raised; it absorbed the nucleonic portion of cosmic rays as well as secondary β- and γ- and environmental radiation. The hard (muon) component of cosmic rays was not filtered by the iron absorber. To prevent muon signals in the sample counter, it was equipped with a ring of fifty screen counters which detected the muons and were run in anticoincidence with the sample counter. The screen counters were even placed inside the sample counter in order to also discriminate against electrons released by cosmic γ-rays from the chamber wall.

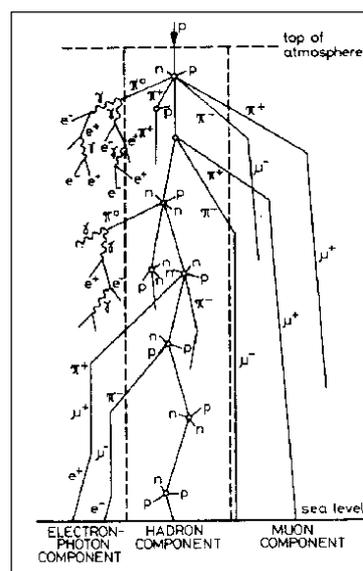

Fig. 10: Schematic representation of cascades caused by primary cosmic rays in the atmosphere

After year-long development of the counter system, Felber and Vychytil carried out methodical investigations regarding the function of the components. Routine dating began in 1965. Stability of electronics and counter (2.4 l, methane at 760 torr/15°) was checked during measurement once a day ($Mn_\alpha$ x-rays following electron capture in $^{55}Fe$). Energy discrimination was not optimized as the neutron generator using the (d,t) reaction was run in the same building in which the samples were prepared. The lower discriminator was set above tritium maximum energy at 22 keV, the upper one at 120 keV, the highest possible energy absorption of $^{14}C$ β-particles in the special counter.[19]

Felber had to install a comprehensively equipped chemical laboratory, too, in which all kinds of samples were pretreated in various procedures to synthesize methane. After careful mechanical cleaning, excavated organic material was pretreated with hot 1% HCl and hot 1% NaOH. The organic sample was burnt in oxygen flux; heated copper oxide completed combustion. Any excess $O_2$ was bound on heated copper. $CO_2$, cleaned by acidified potassium permanganate solution, was frozen out by liquid nitrogen. Shells, after mechanical cleaning, were pretreated with HCl to remove the surface; $CO_2$ was liberated by $H_3PO_4$. The $CO_2$ released into an apparatus for methane synthesis was mixed with a small excess of commercial $H_2$ (because of the above discriminator setting, no special tritium free hydrogen was selected), and circulated over heated (420°C) ruthenium finely divided on aluminum oxide pellets. The synthesized methane was dried in a dry-ice ethanol cooled trap, frozen into a small steel bomb, freed from excess hydrogen by pumping, and stored for four weeks for the purpose of radon decay. Purity of methane was gas chromatographically checked.[19]

Here is a survey of disciplines with which collaboration existed:[20] archeology, pre- and early history, mining, geography, glaciology, climatology, geology, hydro geology, limnology, speleology, sedimentology, mineralogy, petrography, paleontology, botany, palynology, science of soils, forestry, and zoology.

Examples of special challenges in the dating activity were described by Felber in retrospect:[21]

i. One of the unanswered questions of the Alpine ice age was the origin of the big Inn valley terrace in North Tyrol. Its 300-m thick sediments contain e. g. a marine sediment, the varved clay, on which approx. 70 m sands and gravel lay, covered by the ground moraine coating of the last big advance of the Wuerm glacier which carries today's soil. Lacking organic material for which one had looked for decades without success, the terrace sediments were considered early Wuerm glacial and presented as such in geographical excursions. A student participating in an excursion was the one to recognize the indication of a plant relic in a varved-clay outcrop. The yield of a few grams only was considered a sanctuary and dated, resulting in an age of 26,800 years. After sensitizing all clay-mining employees, further material was found in various depths which enabled to set up an age scale up to 31,000 years. Including samples from the ground moraine cover of 11,000 and 10,000 years, a chronology of the last big Wuerm glacier advance was created for the first time.

ii. The Federal Monument Office started a project to record neolithic pile-building residues in Austrian lakes before the still recognizable remainders would be damaged altogether by sporting activities and lake navigation. For several years, Felber collaborated with divers of the Federal Monument Office who carried out underwater measurements and mappings of possible relics near the lake shores in order to trace out pile-building residues. The findings needed to be dated to identify the wooden pile fragments brought up as such from neolithic pile buildings whose age is around 4700 years before present.

Heinz Felber retired in 1987 and was succeeded in radiocarbon dating by Edwin Pak. He had been Felber's student and studied in his thesis an extension of the $^{14}$C dating method by means of isotope enrichment in a separation tube.[22] Although for samples exceeding a certain limiting age the remaining $^{14}$C activity cannot be measured in view of inevitable background, there may still a considerable quantity of $^{14}$C be present if a larger amount of sample material is available. In order to enrich $^{14}$C in methane, a thermal diffusion isotope-separation plant consisting of two columns of 2 m length with a heating power of 5 kW was constructed. For the unusually thick heating tubes a special heater had to be developed. With an available sample of 90 g carbon, the upper age limit of the dating equipment could be raised by 25,000 years, with an uncertainty of the date due to the enrichment < 400 years. The challenge consisted mainly in the sufficiently accurate determination of the enrichment factor ("quantitative enrichment"). Pak's dissertation work is considered a feasibility study assuring that the technique of isotope enrichment worked. However, it was never applied in routinely practice, as it would have required larger amounts of sample material and hence much more capacity of sample treatment which was not affordable at the VRI.

Pak carried on conventional $^{14}$C dating until 2004 when the last few scientists and belongings of the Radium Institute were removed from the historical building in Boltzmanngasse.

## 4.2. Radiocarbon measurement by AMS

In 1993, a new director of the VRI was appointed. As of 1987, it had been an institute of the University Vienna only, rather than a hybrid of University and Academy of Sciences; its name was "Institute for Radium Research and Nuclear Physics". It was still located in the historical building in Boltzmanngasse. Walter Kutschera brought the idea of creating a center for accelerator mass spectrometry, based on an AMS system with a 3-MV-Pelletron tandem accelerator built by National Electrostatics Corporation. The AMS center called VERA (Vienna Environmental Research Accelerator) was set up in a former aristocrats' palace under landmark protection on Währingerstraße and opened in 1996. An important design feature of VERA is the capability of accelerating and transporting ions up to the heaviest element, which allows one to perform AMS experiments with any radionuclide or stable nuclide of the entire nuclear chart. The mounting and adjustment of the accelerator system was done by the staff of the institute, supervised by a technical engineer from NEC. Carbon played an important role in the performance and acceptance tests with regard to both tuning of a $^{12}$C beam through the whole system and achieving the required precision of 1% for $^{14}$C/$^{12}$C ratio measurements of modern standard material. Various systematic studies of $^{14}$C measurements were performed which were reported at the 7$^{th}$ International Conference on AMS in Tucson,

AZ, in 1997.[23] Measurement of $^{14}$C by means of AMS is superior over that of its β-radioactivity in that very little sample material is required; thus, even precious objects can be dated. On the long run, $^{14}$C measurements were intended for a variety of projects:

- Dating of material related to early human activity in the Alps (Ice man)
- Dating of objects from China and Eastern Europe
- $^{14}$CO budget at the high-altitude observatory at Sonnblick (3106 m)
- Identification of carbonaceous aerosols
- Dating of land-snail shell carbonate, connected to loess deposits
- Absolute chronology of early civilizations in Central Europe
- Synchronisation of civilizations in the East Mediterranean

For all $^{14}$C AMS measurements, the sample preparation used methods to prepare carbon samples suitable for use in the sputter ion source of VERA.[24] The procedures of pre-treatment and combustion resemble, therefore, those applied in sample preparation for $^{14}$C radioactivity measurement:

In order to remove non-indigenous carbon from the sample by physical and chemical means, the material is cleaned in an ultrasonic bath to remove adherent particles. For samples like charcoal, textiles, peat, etc., the so-called A-B-A (acid-base-acid) method is applied. A bath of diluted hydrochloric acid dissolves any carbonates possibly present due to ground water percolation, diluted sodium hydroxide solution removes humic acids, a final acid wash ensures that $CO_2$ absorbed during the alkaline step is removed. The temperature and duration of the baths as well as degree of dilution of acid and base are adjusted to the sample nature to minimize the dissolution of sample material. Before and after each step, the samples are washed with double-distilled water.[24]

For dating of fossil bones only the organic fraction of the bone can be used since carbonates of the inorganic bone matrix may exchange with the environment during burial times. The inorganic component is, therefore, removed from the ground bone sample by HCl. As for the organic component, the humic acids are dissolved in NaOH, and the bone collagen turned into gelatine from the residue of the alkaline step.[24]

The cleaned and dried sample material is transferred to a quartz tube together with CuO and silver wire as a binder for sulphur and halogens. The tube is evacuated and sealed and heated in a muffle furnace at 900°C for two hours for complete combustion. The resulting $CO_2$ is transformed into solid carbon by catalytic reduction with hydrogen over iron or cobalt as catalyst at 580°C according to the reaction $CO_2 + H_2 \rightarrow C + 2\ H_2O$. The conversion takes place in a graphitisation apparatus which fulfils highest standards to prevent contamination of the samples: construction components of stainless steel, reaction vessels from quartz and Duran glass, oil free vacuum, achieved with a turbo molecular pump, catalyst freed from traces of $CO_2$. The graphitisation is complete when all the $CO_2$ is reduced to elemental carbon which is deposited on the catalyst and only the excess of $H_2$ is present in the gas volume. The carbon together with the catalyst is used as target material for the $^{14}$C measurements. The mixture is pressed into the 1-mm holes of the aluminium target holders with a recess of 0.5 mm. Up to forty targets can be loaded into the sputter ion source of VERA.[24]

In a review of twenty years of VERA in 2017,[25] two important examples of radiocarbon dating were mentioned: One contributed to the questions of the first appearance of anatomically modern humans in Central Europe and consisted in dating small samples from the teeth of a human skull, which was found more than hundred years ago in a cave in Moravia and preserved in the Anthropology Department of the Natural History Museum of Vienna. The VERA $^{14}$C measurements revealed an age of 34,000 years for this object.[25]

As for the synchronization of Eastern-Mediterranean Civilizations in the Second Millennium BC (SCIEM 2000 project), extensive $^{14}$C dating at a site in the Nile Delta was performed, which is crucial to establish an absolute chronology in the Late Bronze Age. A difference of 120 years was found between the time scale established by $^{14}$C dating and by archaeological reasoning, respectively. A similar difference shows up for the date of the famous volcanic eruption of Santorini in the Aegean around 1600 BC. The SCIEM 2000 project is a ten-year collaboration with archaeologists and Egyptologists. These put forth great efforts to resolve this persistent discrepancy, with evidence emerging that the archaeological time scale may need some correction to arrive at a consensus with the $^{14}$C chronology.[25]

[1] Wikipedia, Willard F. Libby, accessed May 2, 2023.
[2] Samuel Ruben, Martin D. Kamen, Radioactive Carbon of Long Half-Life, Phys. Rev. 57 (1940) 549.
[3] S. A. Korff, On the contribution to the ionization at sea-level produced by the neutrons in the cosmic radiation, Journal of the Franklin Institute 230/6 (1940) 777–779.
[4] Albert Gockel, Luftelektrische Beobachtungen bei einer Ballonfahrt, Physik. Zeitschr. 11 (1910) 280.
[5] Viktor F. Hess, Über Beobachtungen der durchdringenden Strahlung bei sieben Freiballonfahrten, Physik. Zeitschr. 13 (1912) 1084.
[6] W. Bothe, W. Kolhörster, Das Wesen der Höhenstrahlung, Z. Phys. 56 (1929) 751–777.
[7] Arthur H. Compton, Nature of Cosmic Rays, Nature 131 (1933) 713–715.
[8] Brigitte Strohmaier, Robert Rosner (Eds.), Marietta Blau – Stars of Disintegration. Biography of a Pioneer of Particle Physics, Ariadne Press, Riverside, California (2006).
[9] Brigitte Strohmaier, Marietta Blau and the photographic method of particle detection, to be published.
[10] Wilhelm Michl, Zur photographischen Wirkung der α-Teilchen, Sitzungsber. Akad. Wiss. Wien, Math. Naturwiss. Kl. IIa 123 (1914) 1955; Mitt. Inst. Radiumf. 68 (1914).
[11] M. Blau, Über die photographische Wirkung von H-Strahlen aus Paraffin und Atomfragmenten, Z. Physik 48 (1928) 751–764.
[12] M. Blau, Quantitative Untersuchung der photographischen Wirkung von α- und H-Partikeln, Sitzungsber. Akad. Wiss. Wien, Math. Naturwiss. Kl. IIa 139 (1930) 327–347; Mitt. Inst. Radiumf. 259 (1930).
[13] Hertha Wambacher, Untersuchung der photographischen Wirkung radioaktiver Strahlungen auf mit Chromsäure und Pinakryptolgelb vorbehandelte Filme und Platten, Sitzungsber. Akad. Wiss. Wien, Math. Naturwiss. Kl. IIa 140 (1931) 271; Mitt. Inst. Radiumf. 274 (1931).
[14] M. Blau, H. Wambacher, Disintegration processes by cosmic rays with the simultaneous emission of several heavy particles, Nature (London) 140 (1937) 585; M. Blau, H. Wambacher, II. Mitteilung über photographische Untersuchungen der schweren Teilchen in der kosmischen Strahlung. Einzelbahnen und Zertrümmerungssterne, Sitzungsber. Akad. Wiss. Wien, Math. Naturwiss. Kl. IIa 146 (1937) 623–641; Mitt. Inst. Radiumf. 409 (1937).
[15] Heinz Felber, Die Sekundärelektronenemission kathodenzerstäubter Silber-Leichtmetallegierungen, Thesis, Univ. of Vienna, 1950.
[16] Heinz Felber, Peter Vychytil, Messanordnung für energiearme β-Strahlung geringer Intensität, speziell zur Altersbestimmung nach der Radiokohlenstoffmethode, Sitzungsber. Österr. Akad. Wiss., Math. Naturwiss. Kl. II 170 (1961) 179; Mitt. Inst. Radiumf. 546 (1961).
[17] O. C. Allkofer, Cosmic rays on earth, Physik-Daten 25-1, Fachinformationszentrum Energie, Physik, Mathematik, Karlsruhe 1984.
[18] The building was decontaminated according to legal requirements of radiation protection as of 1974.
[19] Heinz Felber, Vienna Radium Institute Radiocarbon Dates I, Radiocarbon 12/1 (1970) 298–318.
[20] Heinz Felber, Edwin Pak, Vienna Radium Institute Radiocarbon Dates IV, Radiocarbon 15/2 (1973) 425–434.
[21] Heinz Felber, private communication to Brigitte Strohmaier, 2007.
[22] Heinz Felber, Edwin Pak, Erweiterung der $^{14}$C-Altersbestimmungsmethode durch quantitative Isotopenanreicherung im Trennrohr, Sitzungsber. Österr. Akad. Wiss., Math. Naturwiss. Kl. II 180 (1971) 299; Mitt. Inst. Radiumf. 630 (1971).
[23] W. Kutschera, P. Collon, H. Friedmann, R. Golser, P. Hille, A. Priller, W. Rom, P. Steier, S. Tagesen, A. Wallner, E. Wild, G. Winkler, VERA: A new AMS facility in Vienna, Proc. 7$^{th}$ Int. Conf. on Accelerator Mass Spectrometry, Tucson, AZ, May 20–24, 1996, Nucl. Instr. Meth. in Phys. Res. B123 (1997) 47–50.
[24] E. Wild, W. Rom, W. Weissenbök, $^{14}$C sample preparation for AMS measurement, Progress Report 1995/96, Institut für Radiumforschung und Kernphysik, Vienna 1996.
[25] Robin Golser, Walter Kutschera, Twenty Years of VERA: Toward a Universal Facility for Accelerator Mass Spectrometry, Nuclear Physics News 27/3 (2017) 29–34.